\documentclass[aps,prapplied,reprint,superscriptaddress,amsmath,amssymb,longbibliography]{revtex4-2}
\usepackage{graphicx}
\usepackage{microtype}
\usepackage[hidelinks,unicode=true,breaklinks=true]{hyperref}
\newcommand{\fref}{f_{\mathrm{ref}}}
\newcommand{\Tframe}{T_{\mathrm{frame}}}
\newcommand{\NP}{N_{\mathrm{P}}}
\begin{document}
\title{In-pixel lock-in readout for wide-field nitrogen-vacancy radio-frequency spectrum analysis}
\author{Shihao Ru}
\affiliation{School of Electrical and Electronic Engineering, Nanyang Technological University, Singapore}
\affiliation{Division of Physics and Applied Physics, School of Physical and Mathematical Sciences, Nanyang Technological University, Singapore}
\affiliation{Centre for Quantum Technologies, Singapore}
\author{Jiakai Chen}
\affiliation{Centre for Quantum Technologies, Singapore}
\author{Xiaodan Lyu}
\affiliation{School of Electrical and Electronic Engineering, Nanyang Technological University, Singapore}
\author{Mingyu Hao}
\affiliation{School of Electrical and Electronic Engineering, Nanyang Technological University, Singapore}
\author{Zitong Xu}
\affiliation{School of Electrical and Electronic Engineering, Nanyang Technological University, Singapore}
\affiliation{Centre for Quantum Technologies, Singapore}
\affiliation{Quantum Science and Engineering Centre, Nanyang Technological University, Singapore}
\author{Edwin Hang Tong Teo}
\affiliation{School of Electrical and Electronic Engineering, Nanyang Technological University, Singapore}
\author{Fedor Jelezko}
\affiliation{Institute for Quantum Optics, Ulm University, Ulm, Germany}
\author{Jesus Z\'u\~niga-Perez}
\affiliation{School of Electrical and Electronic Engineering, Nanyang Technological University, Singapore}
\affiliation{Majulab, International Research Laboratory IRL 3654, CNRS, Universit\'e C\^ote d'Azur, Sorbonne Universit\'e, National University of Singapore, Nanyang Technological University, Singapore}
\author{Weibo Gao}
\email{wbgao@ntu.edu.sg}
\affiliation{School of Electrical and Electronic Engineering, Nanyang Technological University, Singapore}
\affiliation{Division of Physics and Applied Physics, School of Physical and Mathematical Sciences, Nanyang Technological University, Singapore}
\affiliation{Centre for Quantum Technologies, Singapore}
\affiliation{Quantum Science and Engineering Centre, Nanyang Technological University, Singapore}

\begin{abstract}
The growing use of radio-frequency (RF) technologies requires compact, room-temperature sensors that can analyze a wide RF band in real time. Here we present a quantum RF spectrum analyzer based on the gradient-encoded electron spin resonance of a nitrogen-vacancy (NV) ensemble in diamond and read out with a $512\times542$-pixel lock-in camera that demodulates the signal in each pixel. On/off keying the signal under test at the camera's reference frequency lets every pixel extract the spin-resonant photoluminescence modulation. Before digitization, each pixel suppresses the static background and low-frequency noise that limit conventional camera readout. Translating a single permanent magnet tunes the encoded band continuously from 10~MHz to 32~GHz. The magnet position establishes a trade-off between instantaneous bandwidth (up to 4~GHz) and frequency resolution (7--22~MHz), with a 5~ms readout time. The wide-field readout integrates all spectral channels in parallel in a room-temperature sensor head and requires no phase synchronization to the RF carrier. The SNR remains above 70 at an applied power as low as $-32.5$~dBm. The magnetic-field configuration can be adapted to different RF bands without changing the pixel-parallel readout.
\end{abstract}

\maketitle

\section{Introduction}\label{sec:intro}

For receivers based on analog-to-digital conversion, intercepting and analyzing rapidly changing radio-frequency (RF) signals over several gigahertz of instantaneous bandwidth is difficult because sampling rate and power consumption limit direct digitization to a few hundred megahertz~\cite{Gupta2009}. Analog alternatives based on microwave photonics~\cite{Berger2016} or spectral hole burning in rare-earth-doped crystals~\cite{Merkel2014} reach tens of gigahertz, although they sacrifice frequency resolution or require cryogenic operation. Rydberg-atom spectrum analyzers provide high sensitivity from dc to 20~GHz, with instantaneous bandwidth limited to the megahertz scale~\cite{Meyer2021}.

Ensembles of nitrogen-vacancy (NV) centers in diamond~\cite{Gruber1997,Doherty2013,Rondin2014,Barry2020,Degen2017,Casola2018,Du2024} enable broadband spectral analysis at room temperature. In a diamond quantum spectrum analyzer~\cite{Chipaux2015,Magaletti2022,Zhou2025}, a static magnetic-field gradient maps the electron spin resonance (ESR) frequency of each NV center onto position through the Zeeman shift. Similar frequency-to-position encoding has been demonstrated in atomic vapors~\cite{Horsley2015}. An incident microwave field excites only the NV centers resonant with its spectral components, producing a spatially varying photoluminescence (PL) contrast under continuous optical excitation. The camera records this contrast in a single exposure to reconstruct the signal spectrum within the encoded band. The first demonstration achieved 600 MHz instantaneous bandwidth with millisecond refresh rates~\cite{Chipaux2015}. Later architectures span tens of gigahertz with improved spectral resolution and dynamic range~\cite{Magaletti2022,Zhou2025}. The frequency range above 20~GHz overlaps bands used by K- and Ka-band satellite links, 5G millimeter-wave communications, and radar~\cite{Kodheli2021,Rappaport2013}. Wide-field NV imaging can map the near fields of microwave circuits~\cite{Shao2016}.

\begin{figure*}[t]
\centering
\includegraphics[width=\textwidth]{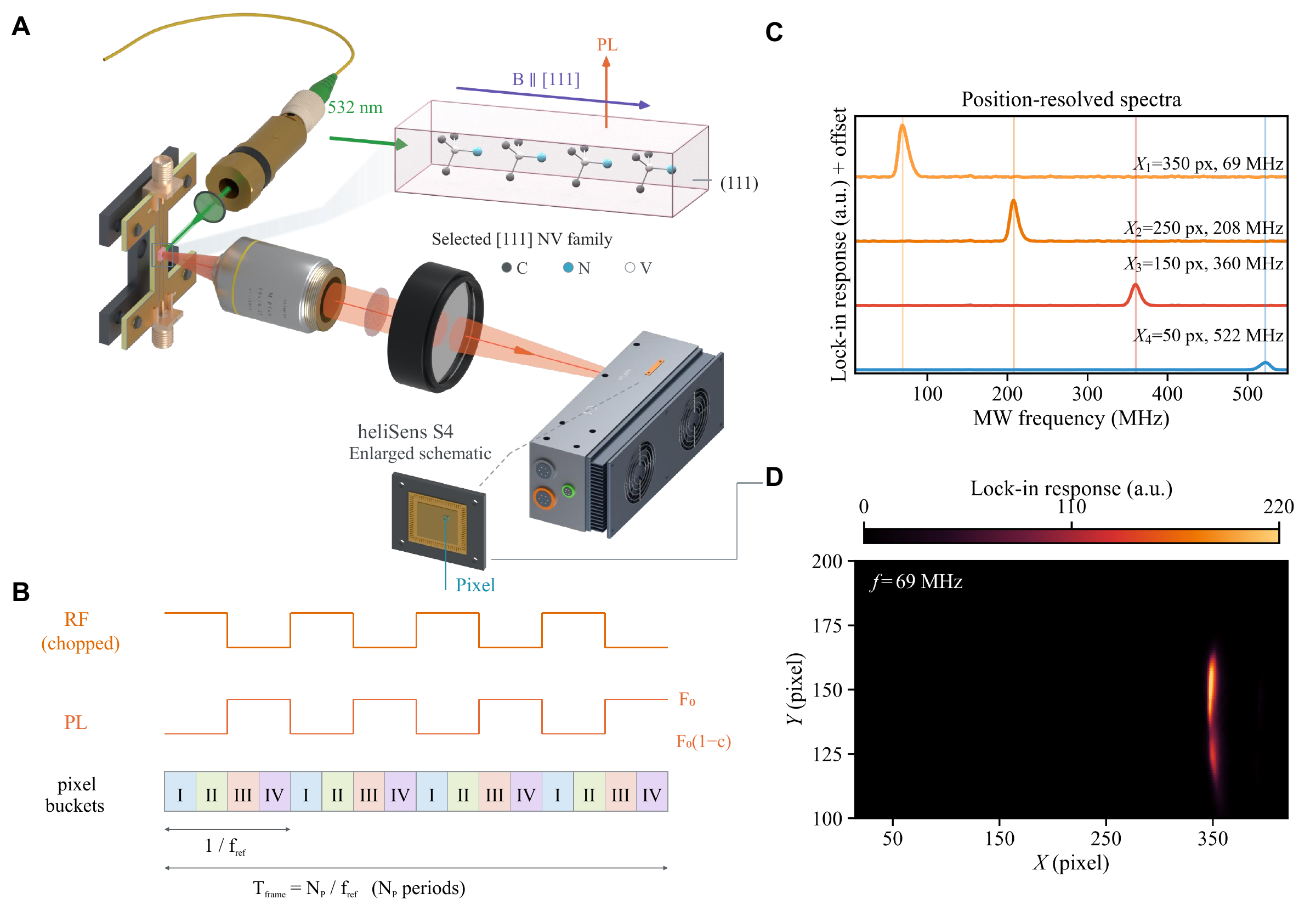}
{
\caption{Wide-field RF spectrum analyzer with in-pixel lock-in detection. (A) Experimental setup. A PCB $\Omega$-loop antenna delivers the chopped RF signal to a (111)-cut diamond. A 532~nm laser illuminates the diamond normal to the (111) surface. An objective collects the PL and images it onto a heliCam~C4 lock-in camera. The bias field is parallel to the selected [111] N--V axis and varies across the sensor. The enlarged views show the selected NV family and a heliSens~S4 pixel. (B) In-pixel demodulation. RF chopping produces a PL square wave with a relative change $c$. Each pixel integrates the photocurrent over four quarter-period windows. After $\NP$ cycles, the pixel returns one complex value for a frame duration $\Tframe = \NP/\fref$. (C) Position-resolved spectra from four pixel columns, vertically offset for clarity. Vertical lines mark the fitted resonance frequencies. (D) Lock-in image at 69~MHz, with the response localized to one pixel column.}
\label{fig:1}
}
\end{figure*}

Optical readout is the main limitation of NV-based microwave spectrum analyzers. The microwave spectrum is encoded as a PL contrast of $10^{-3}$--$10^{-2}$ on a static intensity background. Conventional readout estimates this contrast by subtracting sequentially acquired signal and reference images~\cite{Levine2019,Scholten2021}. The pixel full-well capacity of conventional scientific cameras, typically a few tens of ke$^-$ per exposure, limits this differential imaging scheme~\cite{Wojciechowski2018}. Once the pixels are saturated, additional PL intensity does not improve the single-frame signal-to-noise ratio (SNR). The full-well capacity sets the shot-noise floor. Sequential acquisition leaves residual pump-laser intensity fluctuations and fixed-pattern drift after subtraction. This residual noise eventually dominates photon noise during long-term averaging. Lock-in cameras replace sequential image subtraction with direct in-pixel demodulation. Each pixel separates the modulated PL from the static background before digitization, retains the full photon flux for signal detection, and suppresses low-frequency technical noise. In-pixel demodulation has enabled wide-field NV magnetometry with millisecond and sub-millisecond temporal resolution~\cite{Parashar2022,QDM2023,Webb2022}, but has not yet been applied to NV-based microwave spectrum analysis.

In this work, we use a heliCam~C4 lock-in camera to read out a gradient-encoded NV spectrum analyzer~\cite{Heliotis}. The camera demodulates independently in each of its $512\times542$ pixels. The signal under test is chopped at the camera's reference frequency. Each pixel extracts the resulting spin-resonant PL modulation and suppresses the static background before digitization. Translating a single permanent magnet tunes the encoded band center from 10~MHz to 32~GHz. The analyzer spans up to 4~GHz instantaneously, with 7--22~MHz resolution and a 5~ms readout time. The measured SNR remains above 70 at a driving power of $-32.5$~dBm with a 0.25~s integration time per point. Extrapolating the linear response to $\mathrm{SNR}=1$ gives a minimum detectable antenna input power of $-51$~dBm. A maximum applied drive of $+5$~dBm gives a dynamic range of 56~dB. We systematically vary driving power, pump power, and camera settings to evaluate their effects on sensitivity and resolution. At $\fref = 1$~kHz, multiplicative laser noise limits the SNR. Reconfiguring the magnetic field extends the analyzer to different RF bands without modifying the pixel-parallel readout.

\section{Experimental setup and data analysis}\label{sec:methods}
\label{sec:experiment}

The sensor is a $0.30\times0.88\times1.07$~mm$^3$ (111)-cut HPHT diamond (Viqium, Shanghai) with an NV density of 7.64~ppm. Its polished top surface contains the [111] NV family. The diamond is mounted on a printed circuit board with a 50~$\Omega$-matched $\Omega$-loop antenna. A cylindrical NdFeB magnet provides the bias field and gradient along the in-plane [111] axis. The magnet--diamond distance selects the encoded frequency band. The residual transverse field in this geometry preserves the ODMR contrast~\cite{Zhou2025}.

A continuous-wave 532~nm laser (Sprout-H-10W DPSS) passes through a laser-line filter (Thorlabs FL532-10) and illuminates the diamond from the side, perpendicular to the (111) face. A $10\times$ Olympus objective and a long-pass filter collect the 637--800~nm PL onto the heliCam~C4 ($512\times542$ pixels, external-reference mode).

A Swabian Instruments PulseStreamer~8/2 provides the camera reference $\fref$, the recording trigger, and the 50\%-duty square wave for RF chopping. This common clock synchronizes the chopping and demodulation. A Ceyear~1466L-y signal generator provides the signal under test and accepts the square wave at its pulse-modulation input. The keyed signal passes through a power amplifier and then the loop antenna. We use a Mini-Circuits ZVA-183WA-S+ for the lower bands and a Talent Microwave TLPA18G40G-33-33 for the higher bands.

In a receiver, an RF switch after the antenna would provide chopping without source access. We key the source here for convenience and stream the per-pixel I/Q frames to the acquisition computer.

Channel SNR uses the background-subtracted peak and the off-resonant baseline standard deviation at 0.25~s per frequency point. Gaussian fits with a linear baseline provide the values and uncertainties. The Gaussian line shape describes the inhomogeneous $^{14}$N hyperfine envelope broadened by the field gradient across one channel.

\begin{figure*}[t]
\centering
\includegraphics[width=\textwidth]{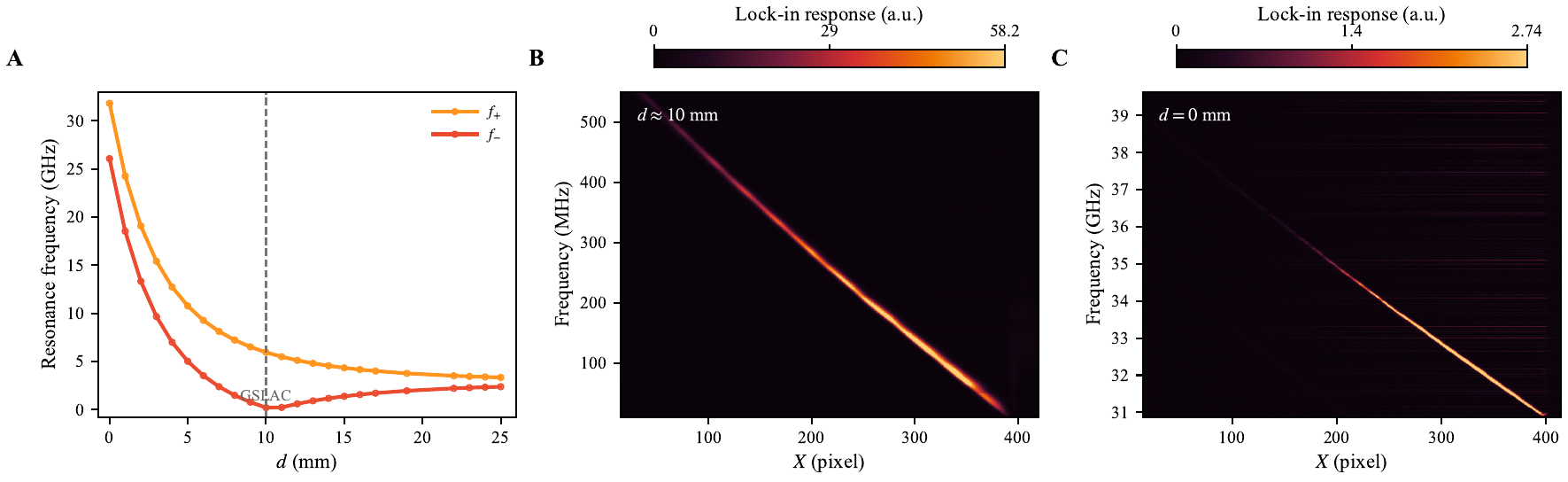}
{
\caption{Frequency tuning at two magnetic-field settings. (A) Measured resonance frequencies $\nu_\pm$ versus magnet--diamond distance $d$. (B) Frequency--position response at $d \approx 10$~mm, near the GSLAC, over 10--550~MHz. (C) Frequency--position response at $d=0$, with an instantaneous bandwidth of 4~GHz.
}
\label{fig:2}
}
\end{figure*}

\section{Frequency-to-position encoding}
\label{sec:result}

{
The analyzer uses a (111)-cut diamond mounted on a PCB $\Omega$-loop antenna that delivers the signal under test. A cylindrical permanent magnet aligns the bias field with the [111] N--V axis and produces a field gradient across the sensor [Fig.~\ref{fig:1}(A)]. A 532~nm laser illuminates the diamond normal to the (111) surface, and the photoluminescence is imaged onto a heliCam~C4 lock-in camera. The reference frequency $\fref$ controls both signal chopping and in-pixel demodulator clocking. Before charge digitization, each pixel extracts the resonant PL modulation and suppresses the static background [Fig.~\ref{fig:1}(B)]. Experimental details are provided in Sec.~\ref{sec:methods}. \par}

Frequency-to-position encoding uses the magnetic-field dependence of the negatively charged NV center's ground-state spin transitions. The zero-field splitting is $D \simeq 2.87$~GHz. For a static field $B_{\mathrm{NV}}$ along the N--V axis, the two ground-state transition frequencies follow
\begin{equation}
  \nu_{\pm} = |D \pm \gamma_e B_{\mathrm{NV}}|,
  \label{eq:zeeman}
\end{equation}
where $\gamma_e \simeq 28.02$~MHz/mT~\cite{Doherty2013}. The bias field $B_{\mathrm{NV}}(x)$ varies monotonically with the imaging coordinate $x$. We use the $\nu_-$ branch ($|0\rangle \to |{-}1\rangle$). The resonance frequency at pixel coordinate $x$ is
\begin{equation}
  \nu(x) = \left| D - \gamma_e B_{\mathrm{NV}}(x) \right| .
  \label{eq:encoding}
\end{equation}
Each spectral component produces a Lorentzian PL reduction in its resonant pixel column, with full width $\Delta\nu$~\cite{Dreau2011}.

The (111) cut aligns the magnetic field and its gradient with the N--V axis across the field of view without fine magnet adjustment~\cite{Zhou2025,Michl2014,Lesik2014}. Magnet position alone selects the operating band. The measurements were performed on the $\nu_-$ branch near the ground-state level anticrossing (GSLAC)~\cite{Broadway2016,Ivady2021}. In this regime, $\gamma_e B_{\mathrm{NV}} \simeq D$ lowers the resonance to a few hundred MHz while the driven transition remains $|0\rangle \to |-1\rangle$. Translating the magnet shifts the encoded band to higher frequencies without changing the readout [Fig.~\ref{fig:2}].

For a continuous-wave signal under test, the analyzer applies on-off keying at the reference frequency $\fref$. The modulation switches the PL intensity between $F_0$ and $F_0[1-c(x)]$, where $c(x)$ is the relative PL change. Each smart pixel integrates the photocurrent over four quarter-period windows per reference cycle. The accumulated signals give the complex amplitude $Z=I+iQ$. One value is returned per frame, with $\Tframe=\NP/\fref$ [Fig.~\ref{fig:1}(B)]. In-pixel AC coupling suppresses the static background before demodulated-signal accumulation. This preserves the output dynamic range for the modulated component. The total detected PL intensity still sets the photon shot-noise floor. The detection band shifts from baseband to $\fref$.

{
We calibrate the frequency-to-position mapping by sweeping the microwave test frequency across the encoded band and recording the demodulated response of every column. Each column responds within its resonance bandwidth [Fig.~\ref{fig:1}(C)]. At 69~MHz, the response appears as a narrow, nearly vertical stripe spanning approximately 50 illuminated rows [Fig.~\ref{fig:1}(D)]. Its horizontal FWHM is approximately nine pixel columns, corresponding to 12~MHz at the local gradient of 1.4~MHz/pixel. A total of 135 pixels lie above half maximum. We bin pixels with the same $x$ coordinate along the rows to form one spectral channel. The number of binned pixels defines $N_{\mathrm{pix}}$ in the noise model (Appendix~\ref{sec:noise}). The stripe curves by approximately three pixels across the illuminated height. This curvature is smaller than the nine-pixel FWHM and bounds the resolution loss from row binning.
\par}

{
Figure~\ref{fig:2}(B) maps $\nu(x)$ across 10--550~MHz. The response is monotonic, with a local slope that changes as the cylindrical magnet's stray field decays with distance. We describe this variation by the local gradient
\par}
\begin{equation}
  g(x) \equiv \left| \frac{d\nu}{dx} \right|
            = \gamma_e \left| \frac{dB_{\mathrm{NV}}}{dx} \right| .
  \label{eq:localgrad}
\end{equation}
{
The gradient ranges from 1.3 to 1.7~MHz/pixel across the field of view. The usable spatial aperture $L$ is the largest contiguous set of columns whose peak response $A_{\max}(x)$ exceeds $10\%$ of $P_{95}[A_{\max}(x)]$. Integrating the local gradient across this aperture gives $\mathrm{BW} = \int_0^L g(x)\,dx = 536$~MHz ($\approx 540$~MHz) [Fig.~\ref{fig:2}(B)]. The fitted band edges are 16 and 552~MHz, with the upper edge 2~MHz beyond the measured scan range.
\par}

\begin{figure}[!t]
\centering
\includegraphics[width=0.94\columnwidth]{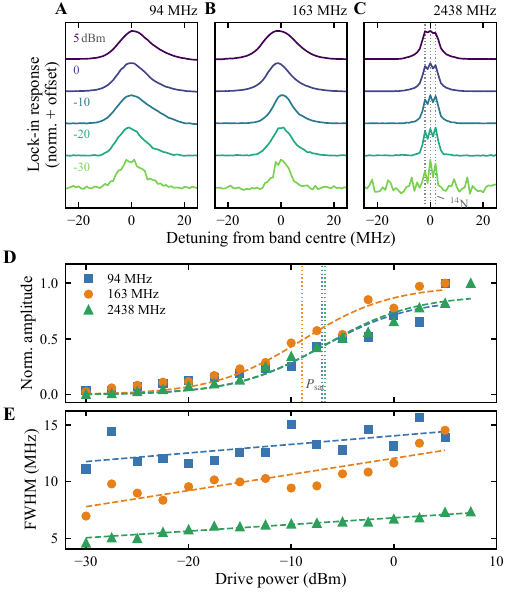}
\caption{Driving-power dependence of signal amplitude and linewidth. All powers refer to the antenna input after amplification. (A)--(C) Peak-normalized, baseline-corrected spectra at 94, 163, and 2438~MHz for different driving powers, vertically offset over a common $\pm25$~MHz detuning range. At low drive, panel (C) resolves the $^{14}$N hyperfine triplet marked by dotted lines. (D) Peak amplitude versus driving power for the three bands, each normalized to its value at the highest power. Saturation fits give $P_{\mathrm{sat}} = -7.0$, $-8.9$, and $-6.7$~dBm for the 94, 163, and 2438~MHz bands, respectively. Vertical dotted lines mark the fitted $P_{\mathrm{sat}}$. (E) Gaussian FWHM versus driving power, with linear trends shown for the low-field (2438~MHz) and near-GSLAC (94 and 163~MHz) measurements.
}
\label{fig:3}
\end{figure}

The frequency resolution at position $x$ is
\begin{equation}
  \delta\nu(x) = \max\!\left(\Delta\nu,\; g(x)\, \delta x\right) .
  \label{eq:resolution}
\end{equation}
Here $\Delta\nu$ is the ODMR linewidth set by the driving power (Fig.~\ref{fig:3}), and $g(x)\delta x$ is the optical resolution expressed as a frequency interval. Integrating the number of resolvable intervals across the aperture gives $N_{\mathrm{ch}} = \int_0^L g(x)/\delta\nu(x)\,dx \approx 49$ independent spectral channels.

{
The magnet--diamond gap $d$ controls the operating frequency. We measured the two $|0\rangle \to |\pm1\rangle$ resonances, $\nu_\pm$, as a function of $d$ [Fig.~\ref{fig:2}(A)], and both branches follow Eq.~(\ref{eq:zeeman}). Reducing the gap increases $B_{\mathrm{NV}}$, raising $\nu_+$ from $D$ to 31.8~GHz at contact while $\nu_-$ reaches zero at the GSLAC before increasing again. A 25~mm magnet translation provides continuous coverage from near zero at the GSLAC to approximately 32~GHz without changing the optics, antenna, or camera.
\par}

{
At contact, the calibration map spans 31--39~GHz across the sensor [Fig.~\ref{fig:2}(C)]. The usable instantaneous bandwidth is 4~GHz. The local gradient of approximately 22~MHz/pixel makes the spatial term $g(x)\delta x$ dominate Eq.~(\ref{eq:resolution}) and set the frequency resolution. The lower-gradient configuration in Fig.~\ref{fig:2}(B) is limited by the ODMR linewidth $\Delta\nu$.

The larger of $\Delta\nu$ and $g(x)\delta x$ sets the frequency resolution [Eq.~(\ref{eq:resolution})]. Lower RF drive narrows $\Delta\nu$ but reduces the SNR in the linewidth-limited regime. Better optical resolution reduces $g(x)\delta x$ in the gradient-limited regime. A larger usable spatial aperture increases the instantaneous bandwidth at the same field gradient.
\par}

\section{Sensitivity and resolution trade-offs}
\label{sec:tradeoffs}

Both the SNR and linewidth depend on the driving power. The demodulated amplitude increases toward saturation according to $A(P) = A_{\max}P/(P+P_{\mathrm{sat}})$. The fitted $P_{\mathrm{sat}}$ values for the three bands differ by no more than 2.3~dB [Fig.~\ref{fig:3}(D)]. For $P \ll P_{\mathrm{sat}}$, extrapolating the linear response to $\mathrm{SNR}=1$ gives a detection limit of $-51$~dBm at the antenna input, with a 56~dB dynamic range extending to $+5$~dBm (Appendix~\ref{sec:noise}, Eq.~(\ref{eq:pminlin})). The inferred detection limits agree within 2~dB across all pump powers.

Increasing the drive raises the signal intensity and broadens the ODMR linewidth. The 2438~MHz linewidth grows from 4.98 to 7.33~MHz between $-25$ and $+7.5$~dBm, and the 163~MHz linewidth increases faster, from 6.96 to 14.5~MHz [Fig.~\ref{fig:3}(E)]. At low drive, the $^{14}$N hyperfine triplet is resolved [Fig.~\ref{fig:3}(C)]. Its 2.16~MHz splitting sets the minimum measured linewidth~\cite{Doherty2013}.

\begin{figure*}[t]
\centering
\includegraphics[width=\textwidth]{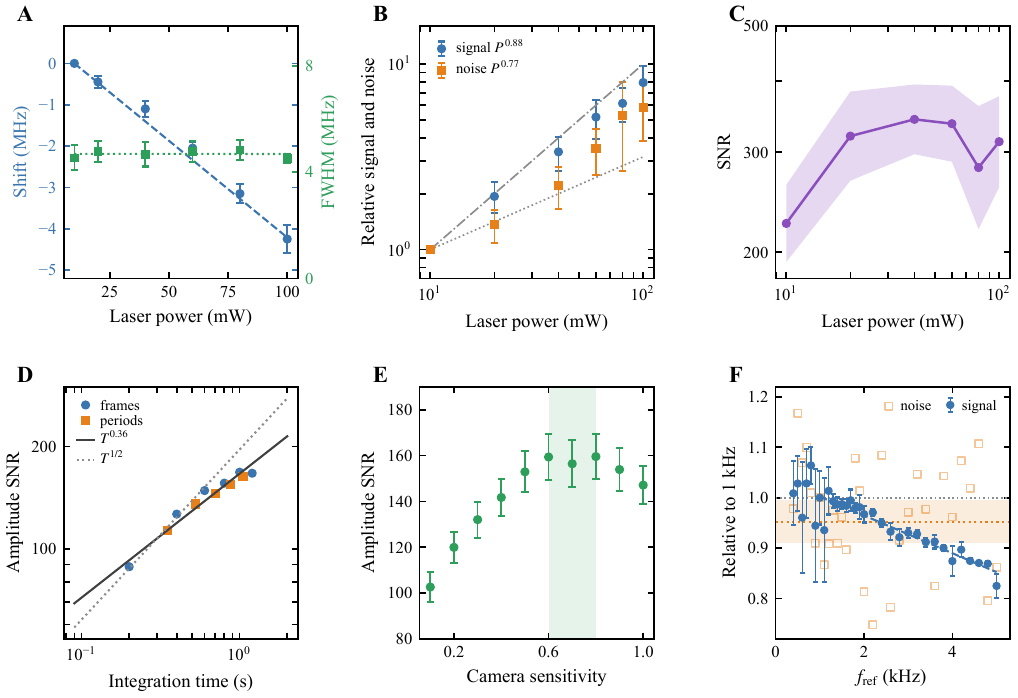}
{
\caption{Effects of pump power and acquisition settings on resonance stability and SNR. (A) Resonance-frequency shift (circles, left axis) and ODMR linewidth (squares, right axis) versus pump power. The measurements use a $3\times37$-pixel spectral channel at $\fref=1$~kHz. Linear fits give slopes of $-46.8$~kHz/mW for the resonance frequency and $-0.0 \pm 2.2$~kHz/mW for the linewidth. (B) Signal and noise versus pump power, normalized to their 10~mW values. Power-law fits give exponents of 0.88 and 0.77, respectively. (C) Geometric-mean SNR versus pump power. The shaded band shows the standard error. (D) Amplitude SNR versus integration time for different numbers of integration periods and burst frames. The fitted SNR scales as $T^{0.364\pm0.011}$. The dotted line shows the $T^{1/2}$ dependence expected for white-noise averaging. (E) Mean SNR versus camera sensitivity. The shaded region marks the 0.6--0.8 plateau, and the error bars show 95\% confidence intervals. (F) Demodulated signal and sideband noise versus $\fref$, normalized to their 1~kHz values. The dashed line fits the signal above 1~kHz. The dotted line and shaded band show the frequency-independent mean noise and its 95\% confidence interval.
}
\label{fig:4}
}
\end{figure*}

Figure~\ref{fig:4} shows how pump power and acquisition settings affect the resonance and SNR. Raising the 532~nm laser power from 10 to 100~mW at fixed camera settings produces little change in the ODMR linewidth, with a fitted slope of $-0.0 \pm 2.2$~kHz/mW. The resonance center shifts by $-4.2$~MHz with a slope of $-46.8$~kHz/mW [Fig.~\ref{fig:4}(A)]. Repeating the fit over 50 pixel columns changes the slope by less than 1\%, indicating that the frequency shift is uniform across the measured region. We estimate laser-induced heating from the resonance shift. Using $\mathrm{d}D/\mathrm{d}T = -74$~kHz/K~\cite{Doherty2013}, the measured shift corresponds to a temperature rise of 0.63~K/mW, or about 57~K from 10 to 100~mW. The pump-induced frequency shift exceeds the $^{14}$N splitting and alters the $\nu(x)$ calibration.

Pump power increases the signal and noise at comparable rates. Figure~\ref{fig:4}(B) gives power-law exponents of $0.88 \pm 0.04$ for the signal and $0.77 \pm 0.05$ for the noise, consistent with multiplicative laser noise. We measure an SNR scaling of $P^{0.10 \pm 0.05}$, compared with the $P^{1/2}$ dependence expected for photon shot noise. Figure~\ref{fig:4}(C) shows that the SNR increases from 225 at 10~mW to 320 at 20~mW and remains unchanged at higher powers. At the $\fref = 1$~kHz operating point, pump noise limits the SNR because the detection band lies within the laser's $1/f$ noise spectrum and includes an appreciable contribution from $\sigma_C^{\mathrm{RIN}}$ (Appendix~\ref{sec:noise}). We use a pump power of 40~mW to maintain the SNR while limiting the thermally induced calibration shift. Raising $\fref$ moves the detection band to a higher frequency, where the $1/f$ noise is lower.

The SNR depends on the total integration time and is insensitive to how that time is divided between integration periods and burst frames. The fitted SNR scales as $T^{0.364 \pm 0.011}$ [Fig.~\ref{fig:4}(D)], below the $T^{1/2}$ expected for white-noise averaging and indicating that part of the noise persists with longer integration. The SNR plateaus at camera sensitivities of 0.6--0.8 and decreases by 7.8\% at 1.0 [Fig.~\ref{fig:4}(E)].

The reference frequency $\fref$ sets the detection band within the noise spectrum. Over the 400~Hz--5~kHz range, the demodulated response peaks near 800~Hz and decreases by 17.5\% between 1 and 5~kHz. The sideband noise shows no resolvable frequency dependence [Fig.~\ref{fig:4}(F)]. The fitted SNR decreases by 17.3\% (Fig.~\ref{fig:supp_controls}(B)). This behavior may reflect a pump $1/f$ knee above the accessible range or a reduction in the chopped ensemble response associated with the finite spin-repolarization time between chops~\cite{Robledo2011}. The linewidth varies by at most 6.2\% across all camera-setting combinations and by only 0.2\% between 1 and 5~kHz, with little effect on spectral resolution.

Conventional wide-field NV spectrum analyzers obtain the ODMR response from differences between sequential images~\cite{Chipaux2015,Magaletti2022,Zhou2025}. Their performance is constrained by the pixel full-well capacity and intensity drift between frames. Our camera demodulates the PL within each pixel before digitization, suppressing the static background and reading the encoded frequency channels in parallel. The pump-power and averaging measurements in Fig.~\ref{fig:4} identify laser-intensity noise and the NV spin response as the main remaining limits.

\section{Conclusions}\label{sec:conclusion}

The analyzer records the gradient-encoded antenna-input spectrum in a single 5~ms camera frame without phase synchronization to the RF carrier. In-pixel lock-in demodulation separates the modulated signal from the static background before digitization, giving a frequency resolution of 7--22~MHz and an SNR above 70 at the weakest applied drive of $-32.5$~dBm. Table~\ref{tab:comparison} compares these figures of merit with representative broadband RF spectrum analyzers. Differences in sensitivity definitions and acquisition protocols prevent a direct performance ranking of the analyzers in Table~\ref{tab:comparison}. Translating the magnet shifts the operating point from the linewidth-limited GSLAC region to a gradient-limited band above 31~GHz. At the high-frequency setting, the usable instantaneous bandwidth reaches 4~GHz, and the tuning range extends to 32~GHz. A stronger bias field would extend the NV transition frequency beyond 32~GHz. For a selected operating band, a frequency-matched antenna could improve microwave delivery and reduce the required input power, provided that its response remains sufficiently flat across the encoded bandwidth.

Laser noise currently limits further improvement in sensitivity. The SNR scales as $T^{0.36}$, plateaus above 20~mW pump power, and decreases as $\fref$ increases from 1 to 5~kHz. These observations are consistent with multiplicative pump noise at the $\fref = 1$~kHz operating point. Active intensity stabilization or a pump with a lower $1/f$ knee could bring the measurement closer to the shot-noise floor. Pump-power stabilization would also limit calibration drift because changing the delivered power shifts the resonance by as much as $-4.2$~MHz over the measured range.

Integrating the lock-in sensor and readout electronics on a single camera board could enable a more compact receiver while retaining pixel-parallel demodulation. Moving the RF chopping to the receiver side would remove the need for transmitter access and allow the analyzer to measure external signals. It could then track multitone and frequency-hopping signals in the K and lower-Ka bands on millisecond timescales. Replacing the magnetic-field gradient with a uniform bias field would provide an imaging mode for mapping the two-dimensional microwave near fields of circuits.

\begin{acknowledgments}
This work is supported by ASTAR (M24M8b0004), Singapore National Research Foundation (NRF-CRP30-2023-0003, NRF-CRP31-0001, NRF2023-ITC004-001 and NRF-MSG-2023-0002) and Singapore Ministry of Education Tier 2 Grant (MOE-T2EP50222-0018).
\end{acknowledgments}

\section*{Author contributions}
S.R. and W.G. conceived the study. S.R. and J.C. designed and built the experimental setup and performed the measurements. S.R. analyzed the data, prepared the figures, and wrote the original draft. J.C., M.H., and Z.X. contributed to the experimental work and validation. X.L. contributed to data analysis and manuscript revision. E.H.T.T., F.J., and J.Z.-P. contributed to the discussion and interpretation of the results and revised the manuscript. W.G. supervised the project. All authors discussed the results and approved the final manuscript.

\section*{Competing interests}
The authors declare no competing interests.

\section*{Data availability}
The data underlying the figures will be deposited in a public repository before publication. The complete raw camera datasets are available from the corresponding author upon reasonable request.

\section*{Code availability}
The analysis code will be deposited in a public repository before publication. The experimental control and acquisition code is available at \url{https://github.com/harleyru/wide_field_lockin}.

\appendix

\section{In-pixel lock-in detection}
\label{sec:lockin}

The analyzer chops the signal under test $s_{\mathrm{RF}}(t)$ at the camera reference frequency $\fref$ with a 50\% duty cycle for synchronous detection (Figs.~\ref{fig:1}(A) and \ref{fig:1}(B)). The PL rate at position $x$ alternates between $F_0$ when the signal is off and $F_0[1 - c(x)]$ when it is on. The total relative PL change is
\begin{equation}
  c(x) = \sum_k C\!\left(P_k\right)
         \mathcal{L}\!\left[\nu_k - \nu(x)\right] ,
  \label{eq:cmap}
\end{equation}
where $C(P_k)$ is the response to a spectral component of power $P_k$, and $\mathcal{L}[\nu_k - \nu(x)]$ is the local ODMR line shape. Chopping maps the RF spectrum onto a PL square wave at $\fref$.

Each camera pixel integrates the photocurrent over four quarter-period windows per reference cycle, giving samples $q_{\mathrm{I}} \ldots q_{\mathrm{IV}}$. Over $\NP$ cycles, the camera accumulates
\begin{equation}
  Z = \sum_{n=1}^{\NP}
      \left( q_{\mathrm{I}} + i q_{\mathrm{II}}
           - q_{\mathrm{III}} - i q_{\mathrm{IV}} \right)_n .
  \label{eq:four_phase}
\end{equation}
One frame lasts $\Tframe = \NP/\fref$. The four-phase demodulation is equivalent to a dual-phase mixer followed by a rectangular low-pass filter. Its transfer function is $H(\Delta f) = \mathrm{sinc}(\Delta f\,\Tframe)$, with an equivalent noise bandwidth of $1/(2\Tframe)$.

For square-wave modulation aligned with the integration windows, Eq.~(\ref{eq:four_phase}) gives $Z=(1+i)cF_0\Tframe/4$. The normalized demodulated amplitude is
\begin{equation}
  \frac{|Z(x)|}{F_0 \Tframe} = g_1\, c(x),
  \qquad g_1 = \frac{\sqrt{2}}{4},
  \label{eq:signal}
\end{equation}
where $F_0$ is the effective detected photon rate after accounting for the camera sensitivity setting $p_s$, which sets the exposed fraction of each quarter period. The factor $g_1$ includes the odd-harmonic response of the square-wave demodulation~\cite{Heliotis}. The signal is divided equally between the two quadratures. A calibrated $45^\circ$ rotation places it on the in-phase axis and leaves the quadrature channel as a real-time noise reference.

In-pixel AC coupling removes the static $F_0$ component before integration~\cite{Heliotis}. The full-well capacity then limits only the modulated signal $\sim cF_0\Tframe$, which is 2--3 orders of magnitude below the background. The available PL sets the detected photon rate $R_0$ per pixel and the shot-noise floor.

\section{Noise model and sensitivity}
\label{sec:noise}

We express each noise source as an equivalent relative PL uncertainty $\sigma_C$ for one spectral channel. A channel sums $N_{\mathrm{pix}}$ pixels assigned to one resonance frequency across the laser stripe. Propagating the Poisson variance through Eq.~(\ref{eq:signal}) gives the shot-noise contribution
\begin{equation}
  \sigma_C^{\mathrm{shot}} =
  \frac{2}{\sqrt{N_{\mathrm{pix}} R_0 T}} ,
  \label{eq:shot}
\end{equation}
for an integration time $T$. Laser intensity noise within the detection band around $\fref$ contributes
\begin{equation}
  \sigma_C^{\mathrm{RIN}} \simeq
  \frac{4}{\pi}\sqrt{\frac{S_{\mathrm{RIN}}(\fref)}{2T}} ,
  \label{eq:rin}
\end{equation}
where $S_{\mathrm{RIN}}(f)$ is the relative intensity noise spectrum. This term is common to all channels.

The shot-noise term decreases as $R_0^{-1/2}$. The RIN term is independent of $R_0$. Across the tenfold pump-power range in Fig.~\ref{fig:4}, $\sigma_C$ follows $P^{-0.10\pm0.05}$ and changes only weakly. This dependence is consistent with the RIN scaling in Eq.~(\ref{eq:rin}). The measurements use $\fref=1$~kHz, where the detection band remains on the $1/f$ slope of the pump laser.

\begin{figure}[!t]
\centering
\includegraphics[width=\columnwidth,trim=0 0 265.53913bp 0,clip]{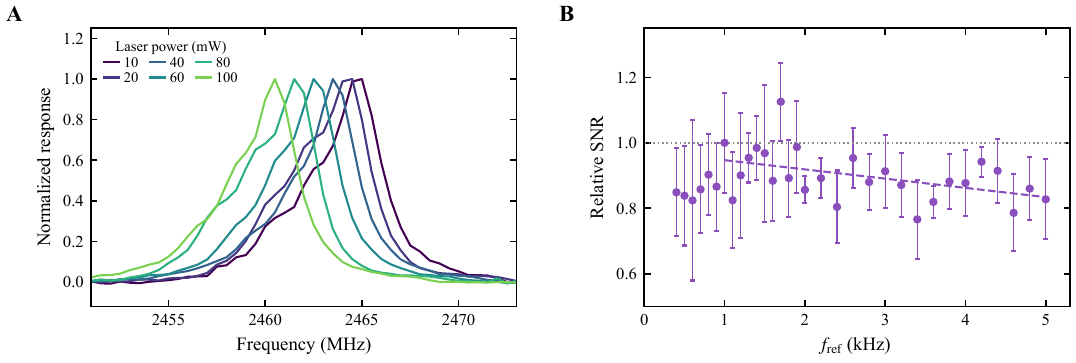}
\par\smallskip
\includegraphics[width=\columnwidth,trim=265.53913bp 0 0 0,clip]{figure5.pdf}
{
\caption{\label{fig:supp_controls}%
Additional pump- and reference-frequency measurements. (A) Peak-normalized ODMR spectra for pump powers from 10 to 100~mW at a fixed driving power of $-20$~dBm and $\fref=1$~kHz. (B) Fitted SNR versus $\fref$, normalized to its value at 1~kHz. The dashed line is a linear fit above 1~kHz, and the error bars show 95\% confidence intervals.}
}
\end{figure}

Raising $\fref$ could move the detection band above the pump's $1/f$ knee. For this apparatus, the demodulated response decreases with $\fref$ [Fig.~\ref{fig:4}(F)]. The sideband noise remains nearly constant. The fitted SNR falls with increasing $\fref$ (Fig.~\ref{fig:supp_controls}(B)). Approaching the shot-noise limit requires a quieter or actively stabilized pump.

A conventional frame-differencing protocol alternates RF-on and RF-off frames of duration $t_f=1/f_{\mathrm{fps}}$ and estimates $c$ from their normalized difference. The full-well capacity limits the photon number per frame to $N_{\mathrm{ph}}\le N_{\mathrm{FW}}$, giving
\begin{equation}
  \sigma_C^{\mathrm{conv}} \gtrsim
  \sqrt{\frac{2}{N_{\mathrm{pix}} \min(R_0 t_f, N_{\mathrm{FW}})}}
  \sqrt{\frac{2 t_f}{T}} .
  \label{eq:conv}
\end{equation}
Differential laser noise adds a term proportional to $\int S_{\mathrm{RIN}}(f)[1-\cos(2\pi f t_f)] \mathrm{sinc}^2(f t_f)\,df$ in the $\lesssim f_{\mathrm{fps}}$ band. Flat-field drift provides another contribution that does not average down. Equation~(\ref{eq:conv}) gives the shot-noise benchmark for frame differencing. A direct comparison requires both protocols to run on the one ensemble under matched conditions with the pump noise reduced to the shot-noise level.

The minimum detectable power per channel follows from the saturation law $C(P)=C_{\max}P/(P+P_{\mathrm{sat}})$ for CW ODMR of a driven two-level transition~\cite{Dreau2011,Chipaux2015}. Fits to the bands in Fig.~\ref{fig:3} give $P_{\mathrm{sat}}=-7$ to $-9$~dBm at the antenna input. In linear power units, inversion gives
\begin{equation}
  P_{\min}(T) = P_{\mathrm{sat}}
  \frac{\sigma_C(T)}{C_{\max} - \sigma_C(T)} ,
  \label{eq:pmin}
\end{equation}
where $\sigma_C(T)$ includes the shot-noise and RIN terms in Eqs.~(\ref{eq:shot}) and~(\ref{eq:rin}).

The reported $-51$~dBm sensitivity is obtained from the measured linear response. More than 20~dB below $P_{\mathrm{sat}}$, the SNR is proportional to $P_{\mathrm{RF}}$. In dBm, extrapolation to unit SNR gives
\begin{equation}
  P_{\min}\,[\mathrm{dBm}] = P_{\mathrm{RF}}\,[\mathrm{dBm}] -
  10\log_{10}(\mathrm{SNR}).
  \label{eq:pminlin}
\end{equation}
The three lowest drive settings, $-32.5$ to $-27.5$~dBm, agree within $\pm1.5$~dB. The six pump powers in Fig.~\ref{fig:4} give $P_{\min}=-50.0$ to $-52.0$~dBm, reported as $-51$~dBm. This value refers to the antenna input and an integration time of 0.25~s per frequency point. The largest applied drive of $+5$~dBm gives a 56~dB dynamic range. Response saturation above $P_{\mathrm{sat}}$ reduces sensitivity near the upper end. The 21-bit-equivalent range restored by in-pixel background suppression~\cite{Heliotis} exceeds this applied range.

\section{Analyzer metrics}
\label{sec:metrics}

\begin{table*}[t]
\centering
\caption{Comparison with representative broadband RF spectrum analyzers. Values are taken from the cited references. Sensitivity definitions differ among platforms. The Rydberg value is a spectral density measured with an RF local oscillator, Ref.~\cite{Chipaux2015} reports applied drive power, and the remaining values are minimum detectable powers. For this work, the sensitivity is extrapolated to unit SNR from SNR $>$ 70 at $-32.5$~dBm and refers to the antenna input with 0.25~s integration per point. Dashes denote unreported quantities. SHB denotes spectral hole burning, RT room temperature, and LO local oscillator.}
\label{tab:comparison}
\footnotesize
\renewcommand{\arraystretch}{1.15}
\setlength{\tabcolsep}{3pt}

\textbf{Frequency characteristics}\\[0.35em]
\begin{tabular}{@{}p{0.25\textwidth}p{0.21\textwidth}p{0.18\textwidth}p{0.18\textwidth}@{}}
\hline
Platform & Coverage & Instant. bandwidth & Resolution \\
\hline
SHB photonic rainbow~\cite{Berger2016} & multi-GHz & $>$20~GHz & 400 channels \\
Rydberg heterodyne~\cite{Meyer2021} & dc--20~GHz & 4~MHz & --- \\
NV, frame differencing~\cite{Chipaux2015} & near 2.87~GHz & 600~MHz & 7~MHz \\
NV, scanned gradient~\cite{Magaletti2022} & 10~MHz--27~GHz & 0.3--4~GHz & 1--50~MHz \\
NV, (111) wide field~\cite{Zhou2025} & 20~MHz--18~GHz & 2~GHz & 5.8~MHz \\
This work, in-pixel lock-in & 10~MHz--32~GHz & up to 4~GHz & 7--22~MHz \\
\hline
\end{tabular}

\vspace{0.8em}
\textbf{Acquisition and operating characteristics}\\[0.35em]
\begin{tabular}{@{}p{0.23\textwidth}p{0.14\textwidth}p{0.22\textwidth}p{0.14\textwidth}p{0.11\textwidth}@{}}
\hline
\raggedright Platform & Spectrum time & Sensitivity & Dynamic range & Operation \\
\hline
\raggedright SHB photonic rainbow~\cite{Berger2016} & $<$0.1~ms & --- & --- & 3~K \\
\raggedright Rydberg heterodyne~\cite{Meyer2021} & --- & $-120$~dBm/Hz & $>$80~dB & RT, RF LO \\
\raggedright NV, frame differencing~\cite{Chipaux2015} & 4~ms & hundreds of~$\mu$W & --- & RT \\
\raggedright NV, scanned gradient~\cite{Magaletti2022} & 2--600~ms & $-17$~dBm & 40~dB & RT \\
\raggedright NV, (111) wide field~\cite{Zhou2025} & --- & $-55$~dBm & 65~dB & RT \\
\raggedright This work, in-pixel lock-in & 5~ms & $-51$~dBm (0.25~s) & 56~dB & RT, no LO \\
\hline
\end{tabular}
\end{table*}

The temporal resolution is one lock-in frame, $\Tframe=\NP/\fref$. The heliCam~C4 supports frame times from 0.25 to 325~ms and full-frame streaming at approximately 5~kHz~\cite{Heliotis}. Most measurements use $\NP=5$ and $\fref=1$~kHz, giving a 5~ms spectrum time. Calibration takes longer because it scans the microwave frequency across the band. After calibration, all channels are acquired in parallel within one frame. Chopping reduces the average received power by 3~dB. The four-phase demodulation adds the factor $g_1=\sqrt{2}/4$ in Eq.~(\ref{eq:signal}).

The magnet assembly sets the tunable range. A single cylindrical magnet covers the GSLAC to approximately 32~GHz on the $|0\rangle\to|+1\rangle$ branch [Fig.~\ref{fig:2}(A)]. Table~\ref{tab:comparison} compares this coverage with Refs.~\cite{Magaletti2022,Zhou2025}. NV ODMR has been measured at comparable fields and frequencies~\cite{Stepanov2015}. The GSLAC sets the low-frequency limit~\cite{Zhou2025}.

\begingroup
\endgroup

\end{document}